\documentclass[twocolumn,english,prl]{revtex4}
\usepackage[T1]{fontenc}
\usepackage[latin9]{inputenc}
\usepackage{color}
\usepackage{textcomp}
\usepackage{amsmath}
\usepackage{amssymb}
\usepackage{graphicx}

\makeatletter

\providecommand{\tabularnewline}{\\}

\@ifundefined{textcolor}{}
{%
 \definecolor{BLACK}{gray}{0}
 \definecolor{WHITE}{gray}{1}
 \definecolor{RED}{rgb}{1,0,0}
 \definecolor{GREEN}{rgb}{0,1,0}
 \definecolor{BLUE}{rgb}{0,0,1}
 \definecolor{CYAN}{cmyk}{1,0,0,0}
 \definecolor{MAGENTA}{cmyk}{0,1,0,0}
 \definecolor{YELLOW}{cmyk}{0,0,1,0}
}

\usepackage{babel}

\usepackage{babel}

\usepackage{babel}

\makeatother

\usepackage{babel}
\begin{document}

\title{What the Timing of Millisecond Pulsars Can Teach us about Their Interior}

\author{Mark G. Alford and Kai Schwenzer}

\address{Department of Physics, Washington University, St. Louis, Missouri,
63130, USA}
\begin{abstract}
The cores of compact stars reach the highest densities in nature and
therefore could consist of novel phases of matter. We demonstrate
via a detailed analysis of pulsar evolution that precise pulsar timing
data can constrain the star's composition, through unstable global
oscillations (r-modes) whose damping is determined by microscopic
properties of the interior. If not efficiently damped, these modes
emit gravitational waves that quickly spin down a millisecond pulsar.
As a first application of this general method, we find that ungapped
interacting quark matter is consistent with both the observed radio
and x-ray data, whereas for ordinary nuclear matter some additional
enhanced damping mechanism is required. 
\end{abstract}
\maketitle
Pulsars are believed to be ultradense compact objects that may consist
of nuclear matter \cite{Lattimer:2004pg} or may include more exotic
material such as quark matter \cite{Itoh:1970uw,Witten:1984rs,Alford:2007xm}.
There is a wealth of very precise radio---and increasingly also high-energy
\cite{TheFermi-LAT:2013ssa}---pulsar timing data \cite{Manchester:2004bp},
showing that they are extremely stable systems with known frequency
and spindown (SD) rate. The goal of this paper is to show that these
data can be used to constrain hypotheses about the interior composition
of the star. Our approach relies on r-modes (RMs) \cite{Andersson:1997xt,Andersson:2000mf},
global oscillations which are unstable via the Friedman-Schutz mechanism
\cite{Friedman:1978hf}. If they are not effectively damped, r-modes
grow spontaneously and emit gravitational waves (GWs), spinning the
star down. Different possible phases of dense matter have different
viscosities, and hence differ in their ability to damp r-modes. Therefore
observations of high-spin pulsars indicate that sufficiently strong
damping must be present, constraining the possible phases of matter
in the star.

The macroscopic state of the star is specified by its angular velocity
$\Omega=2\pi f$, its core temperature $T$, %(which for X-ray pulsars can be estimated by spectral fitting), and
the amplitude $\alpha$ of the r-mode (which is unobservable). The
evolution is determined by conservation equations % between the different components of the system and external radiation fields\cite{Owen:1998xg,Wagoner:2002vr}
which involve energy loss rates, namely the power $P_{G}$ fed into
the r-mode by radiating gravitational waves, the dissipated power
$P_{D}$ that heats the star and the thermal luminosity $L$ that
cools it. If not globally so at least over certain temperature ranges,
they follow power laws which for the $m\!=\!2$ r-mode read 
\begin{equation}
P_{G}=\hat{G}\Omega^{8}\alpha^{2},\; P_{D}=\hat{D}T^{\delta}\lambda^{\Delta}\Omega^{\psi}\alpha^{\phi},\; L=\hat{L}T^{\theta}\lambda^{\Theta},\label{eq:powers}
\end{equation}
where $\phi\!=\!2$ at small $\alpha$, $\lambda\!\equiv\!1+\!\sigma\log\!\left(\Lambda/T\right)$
are logarithmic correction factors that arise from non-Fermi liquid
(NFL) effects in certain forms of quark matter \cite{Schwenzer:2012ga}
and we neglect an amplitude dependence of the luminosity since it
is only relevant for the unrealistic case $\alpha\!=\! O\!\left(1\right)$
\cite{Alford:2012yn}. The prefactors in (\ref{eq:powers}) are given
in tab.~\ref{tab:parameterization}. They are given by a few dimensionless
parameters that encode the relevant properties of the star, see tab.~\ref{tab:parameter-values}.
We will first review and refine constraints on the composition from
measurements of $f$ and $T$, concluding that currently known damping
mechanisms have difficulty explaining the pulsar data within minimal
hadronic matter models of neutron stars, which include only viscous
damping which can be reliably estimated. We then show how this conclusion
is confirmed and enhanced by measurements of $f$ and $\dot{f}$ (``timing
data'').
\begin{table*}
\begin{tabular}{|c|c|c|c|c|c|c|c|c|c|c|c|c|c|c|c|c|}
\hline 
compact star  & $R\left[km\right]$  & $\tilde{I}$  & $\tilde{J}$  & $\tilde{S}$  & $\tilde{V}$  & $\tilde{L}$  & $\delta_{{\rm SV}}$  & $\psi_{{\rm SV}}$  & $\delta_{{\rm BV}}$  & $\psi_{{\rm BV}}$  & $\Delta_{{\rm BV}}$  & $\delta_{{\rm EL}}$  & $\psi_{{\rm EL}}$ & $\theta_{\nu}$  & $\Theta_{\nu}$  & $\theta_{\gamma}$\tabularnewline
\hline 
SS $1.4\, M_{\odot}$ (NFL)  & $11.3$  & $0.374$  & $3.08\!\times\!10^{-2}$  & $3.49\!\times\!10^{-6}$  & $3.53\!\times\!10^{-10}$  & $1.74\!\times\!10^{-6}$  & $-\frac{5}{3}$  & $2$  & $2$  & $4$  & $0$$\left(4\right)$  & - & - & $6$  & $0$$\left(2\right)$  & $4{\color{red}\iota}$\tabularnewline
\hline 
NS $1.4\, M_{\odot}$  & $11.5$  & $0.283$  & $1.81\!\times\!10^{-2}$  & $7.68\!\times\!10^{-5}$  & $1.31\!\times\!10^{-3}$  & $1.91\!\times\!10^{-2}$  & $-\frac{5}{3}$  & $2$  & $6$  & $4$  & $0$  & $\frac{\delta_{{\rm SV}}}{2}$ & $\frac{5}{2}$ & $8$  & $0$  & $4{\color{red}\iota}$\tabularnewline
\hline 
\end{tabular}

\protect\protect\caption{\label{tab:parameter-values}Parameters characterizing the strange
star (SS) and neutron star (NS) considered in this work \cite{Alford:2010fd}.
The exponents $\delta_{{\rm SV}}$, $\cdots$, $\theta_{\gamma}$
($\Delta_{{\rm SV}}$,$\Delta_{{\rm EL}}$,$\Theta_{\gamma}\!=\!0$)
arise in the parameterizations eq.~(\ref{eq:powers}) for mechanisms
in tab.~\ref{tab:parameterization} along with the corresponding
constants $\tilde{I}$, $\cdots$, $\tilde{L}$.}
\end{table*}

\begin{table}
\begin{tabular}{|c||c|}
\hline 
GW luminosity  & $\hat{G}\equiv2^{17}\pi/\left(3^{8}5^{2}\right)\tilde{J}^{2}GM^{2}R^{6}$\tabularnewline
\hline 
\hline 
Shear viscosity (SV) & $\hat{D}=5\tilde{S}\Lambda_{{\rm QCD}}^{3-\delta_{sv}}R^{3}$\tabularnewline
\hline 
Bulk viscosity (BV, low $T$)  & $\hat{D}=2^{3}/\left(3^{3}7\right)\Lambda_{{\rm QCD}}^{9-\delta_{bv}}\tilde{V}R^{7}/\Lambda_{{\rm EW}}^{4}$\tabularnewline
\hline 
Ekman layer (EL)  & $\hat{D}=5\left(\frac{2}{3}\right)^{9/2}\!\frac{3401+2176\sqrt{2}}{11!!}\sqrt{\hat{\eta}_{c}\rho_{c}}R_{c}^{4}$\tabularnewline
\hline 
\hline 
Neutrino luminosity  & $\hat{L}=4\pi R^{3}\Lambda_{{\rm QCD}}^{9-\theta}\tilde{L}/\Lambda_{{\rm EW}}^{4}$\tabularnewline
\hline 
Photon luminosity  & $\hat{L}=\pi^{3}/15\, R^{2}\hat{X}^{4}$\tabularnewline
\hline 
\end{tabular}\protect\protect\caption{\label{tab:parameterization}Parameters in the general parameterization
eq.~(\ref{eq:powers}) for the energy loss rates, in terms of the
star's mass $M$ and radius $R$, the gravitational constant $G$,
generic scales $\Lambda_{{\rm QCD}}$ and $\Lambda_{{\rm EW}}$. For
Ekman damping parameters see \cite{Lindblom:2000gu}. The dimensionless
constants $\tilde{J}$, $\tilde{V},$ $\tilde{S}$ and $\tilde{L}$
encode properties of the stellar interior \cite{Alford:2012yn} and
are evaluated in table \ref{tab:parameter-values}.}
\end{table}

\begin{figure*}
\includegraphics{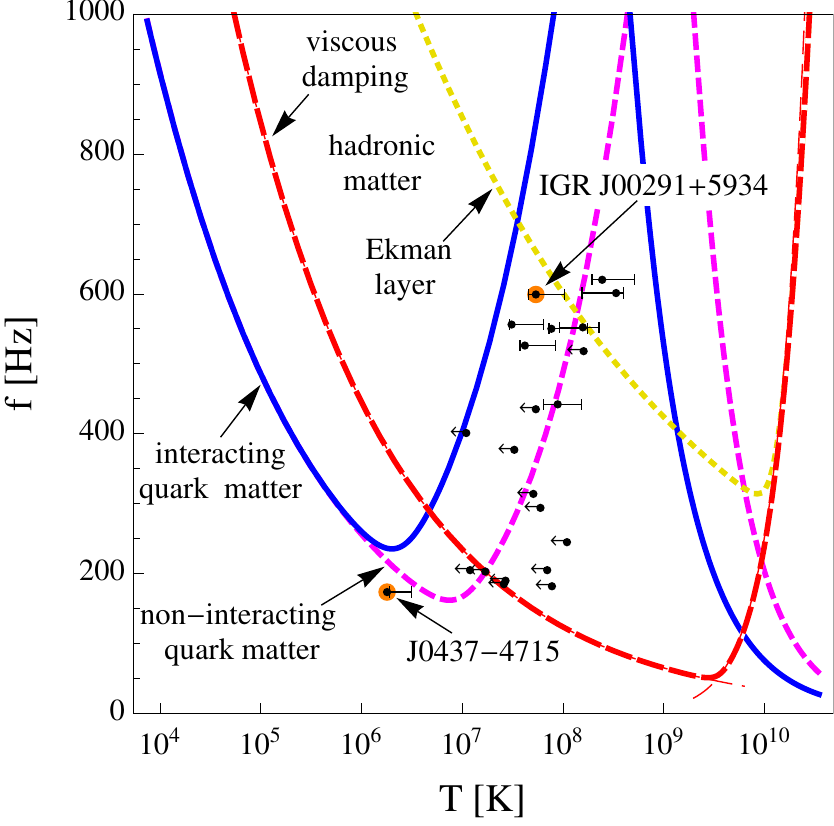} \includegraphics{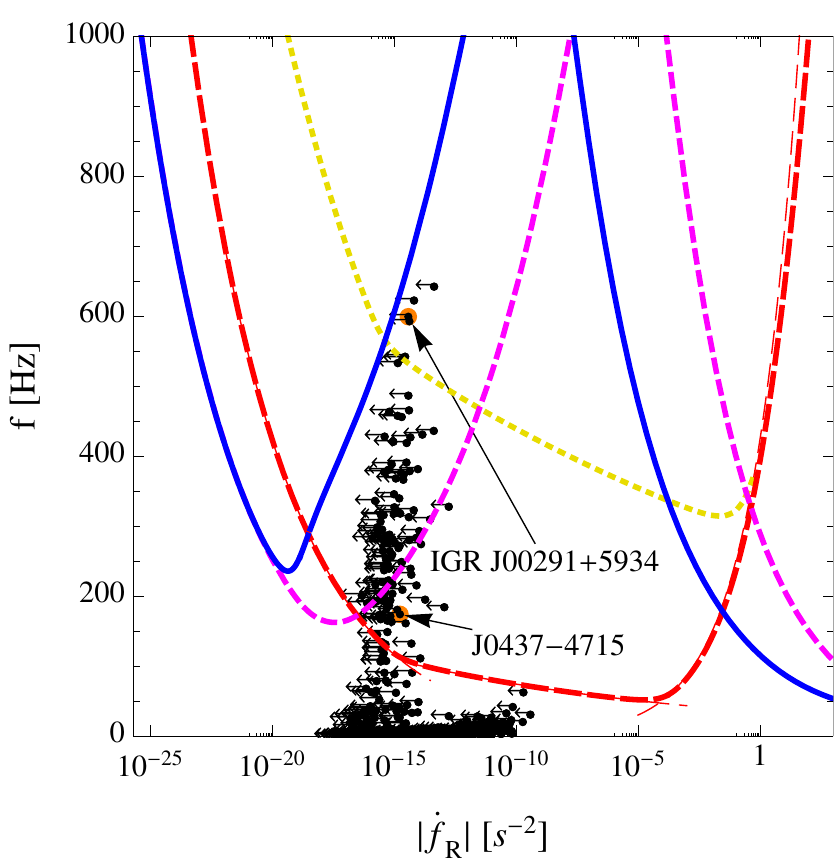}

\protect\protect\caption{\label{fig:exclusion-regions}Boundaries of the r-mode instability
regions for different star compositions compared to pulsar data. \emph{Left:}
Standard static instability boundary compared to x-ray data \cite{Haskell:2012,Tomsick:2004pf}
with error estimates from different envelope models \cite{Gudmundsson:1983ApJ,Potekhin:1997}.
\emph{Right:} Dynamic instability boundary in timing parameter space
compared to radio data \cite{Manchester:2004bp} (all data points
are upper limits for the r-mode component of the spindown). The curves
represent: $1.4\, M_{\odot}$ neutron star with standard viscous damping
\cite{Shternin:2008es,Sawyer:1989dp} (long-dashed) and with additional
boundary layer rubbing \cite{Lindblom:2000gu} at a rigid crust (dotted)
as well as $1.4\, M_{\odot}$ strange star \cite{Madsen:1992sx,Heiselberg:1993cr}
(short-dashed) and same with long-ranged NFL interactions causing
enhanced damping \cite{Schwenzer:2012ga} (using the strong coupling
$\alpha_{s}\!=\!1$) (solid)---more massive stars are not qualitatively
different. The thin curves show for the neutron star exemplarily the
analytic approximation for the individual segments. The encircled
points denote the only ms-radio pulsar J0437\textminus 4715 with a
temperature estimate and the only LMXB IGR J00291+5934 that has been
observed to spin down during quiesence.}
\end{figure*}

The left panel of fig.~\ref{fig:exclusion-regions} shows $T$-$f$
data for low mass x-ray binaries (LMXBs) \cite{Haskell:2012}, which
are being heated and potentially spun up by accretion from a companion.
$T$ is the core temperature, inferred from X-ray spectra using a
model of the envelope \cite{Gudmundsson:1983ApJ}. These involve uncertainties
(estimated by the error-bars) or provide only upper limits (left-pointing
arrows). The figure also shows \emph{static} instability boundaries
\cite{Andersson:2000mf} for a few hypothesized star compositions.
The boundaries are determined by $P_{G}\!=\!\left.P_{D}\right|_{\alpha\to0}$
and explicitly given for the individual segments with a given dominant
damping mechanism by \cite{Alford:2010fd,Reisenegger:2003cq}

\begin{equation}
\Omega_{{\rm IB}}\!\left(T\right)=\left(\hat{D}T^{\delta}\lambda^{\Delta}/\hat{G}\right)^{1/(8-\psi)}\,.\label{eq:temperature-instability}
\end{equation}
The region above a boundary is where dissipation is insufficient to
damp the r-modes. The dissipation arises from shear viscosity \cite{Shternin:2008es,Heiselberg:1993cr},
bulk viscosity \cite{Sawyer:1989dp,Alford:2010gw,Schwenzer:2012ga}
or another mechanism like surface rubbing in a viscous boundary layer
at a solid crust \cite{Lindblom:2000gu}.

The solid line is the instability boundary for a model of interacting
ungapped quark matter \cite{Schwenzer:2012ga} which is compatible
with the data via the \emph{no r-mode} scenario, where r-modes are
completely damped; this is due to the resonant enhancement of bulk
viscosity \cite{Madsen:1998qb,Madsen:1999ci,Schwenzer:2012ga} which
creates a large stability window at $T\sim10^{7}-10^{9}\,{\rm K}$.
In contrast, a model of non-interacting quark matter (short-dashed)
does not explain the data.

The long-dashed line is the instability boundary for stars made of
hadronic matter, taking into account viscous damping only. Most of
the data points lie above this line, indicating that this model would
leave r-modes unsuppressed. Even if we add maximum viscous damping
at the crust-core boundary \cite{Lindblom:2000gu}, requiring an implausibly
thin (cm-size) Ekman layer, using the improved shear viscosity result
\cite{Shternin:2008es} we still get an instability line (dotted)
that is below some points %
\footnote{Magnetic fields can enhance Ekman damping, but have a minor impact
for $B\!<\!10^{9}\,{\rm G}$ in old ms-pulsars \cite{Mendell:2001nz}.%
}. Therefore, the hadronic matter model is only compatible with the
data if there is some additional damping mechanism, or in a \emph{saturated
r-mode} scenario, where non-linear damping $P_{D}\!\left(\alpha\right)$
($\phi>2$) limits r-modes to a tiny amplitude $\alpha_{{\rm sat}}$,
determined by the condition $P_{G}\!\left(\alpha_{{\rm sat}}\right)\!=\! P_{D}\!\left(\alpha_{{\rm sat}}\right)$.
We use a general power-law parametrization of the saturation amplitude
$\alpha_{{\rm sat}}=\hat{\alpha}_{{\rm sat}}T^{\beta}\Omega^{\gamma}$
as realized for proposed mechanisms \cite{Bondarescu:2013xwa,Haskell:2013hja,Lindblom:2000az}.

To see how small $\alpha_{{\rm sat}}$ in hadronic matter has to be,
we need to calculate the spindown evolution. It is crucial to note
that the thermal evolution is always faster than the spindown \cite{Alford:2012yn},
so the temperature reaches a steady state where cooling matches heating
\cite{Bondarescu:2013xwa}, $P_{G}\!=\! L$, giving

\begin{equation}
\Omega_{{\rm SD}}\!\left(T;\hat{\alpha}_{{\rm sat}}\right)=\left(\hat{L}T^{\theta-2\beta}\lambda^{\Theta}/\left(\hat{G}\hat{\alpha}_{{\rm sat}}^{2}\right)\right)^{1/\left(8+2\gamma\right)}\,.\label{eq:steady-state}
\end{equation}

\begin{figure*}
\includegraphics{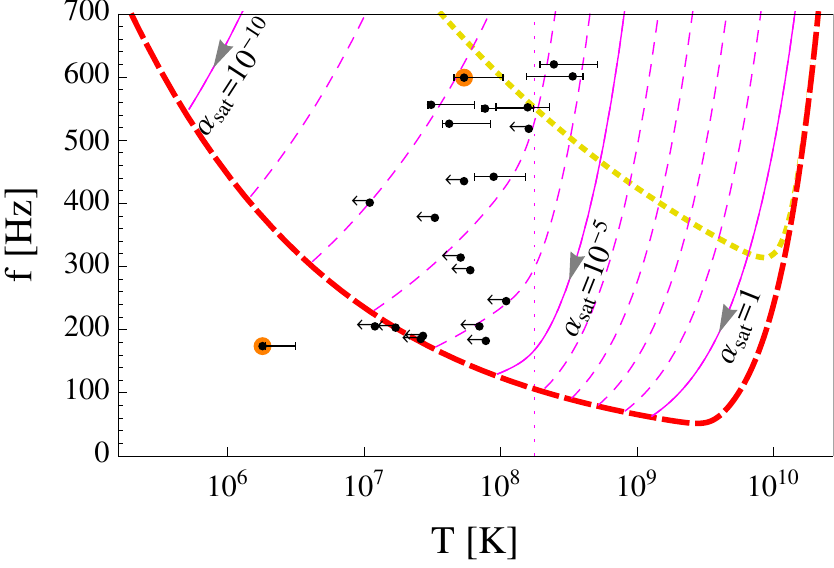} \includegraphics{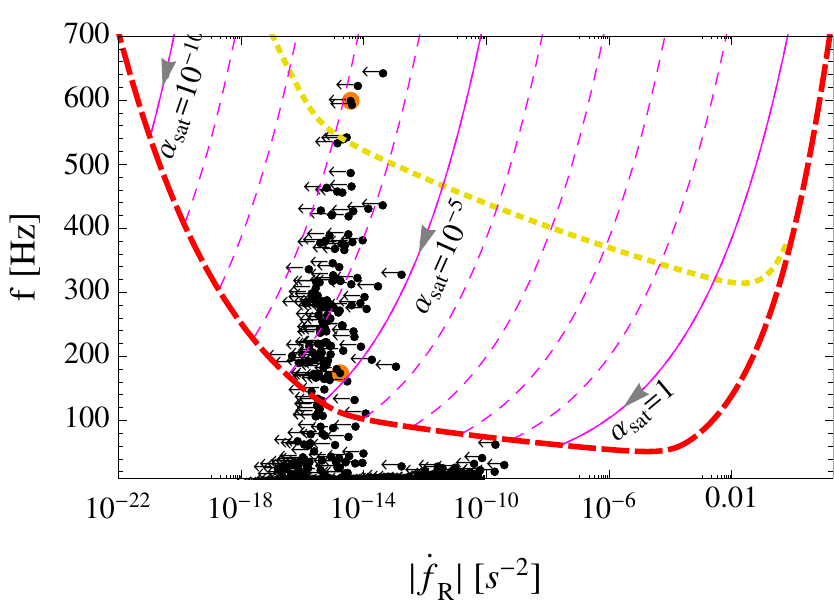}

\protect\protect\caption{\label{fig:steady-state}The thermal steady-state spindown curves,
along which the star evolves, for several $T$- and $\Omega$-independent
saturation amplitudes. Shown are also the boundaries of the instability
regions for neutron stars with different damping sources as discussed
in fig. \ref{fig:exclusion-regions}. \emph{Left:} Static instability
boundary compared to x-ray data \cite{Haskell:2012,Tomsick:2004pf}.
Evolution curves are shown for a $1.4\, M_{\odot}$ neutron star with
modified Urca cooling. The vertical line gives the temperature below
which photon emission replaces neutrino emission as the dominant cooling
mechanism. \emph{Right:} Dynamic instability boundary in timing parameter
space compared to radio data \cite{Manchester:2004bp}.}
\end{figure*}

In Fig.~\ref{fig:steady-state} (left panel) we plot the same LMXB
observations along with the spindown curves eq.~(\ref{eq:steady-state})
for hadronic matter for a range of values of $\alpha_{{\rm sat}}$.
We assume photon and modified Urca cooling, with $\alpha_{{\rm sat}}$
independent of $T$ and $\Omega$ \cite{Owen:1998xg}. For surface
luminosity we take the surface temperature $T_{s}$ to be related
to core temperature $T$ via the unaccreted envelope model \cite{Gudmundsson:1983ApJ},
$T_{s}\!=\!\hat{X}T^{\iota}$, with $\hat{X}\!\approx\!34.6\,{\rm K}^{0.45}g_{s14}^{1/4}$,
$\iota\!\approx\!0.55$. The data points are for LMXBs which are heated
by accretion, so they can lie to the right of the spindown curves.
We conclude that for these sources $\alpha_{{\rm sat}}\lesssim O\!\left(10^{-8}\!-\!10^{-6}\right)$
and similar bounds were obtained in \cite{Mahmoodifar:2013quw}. Moreover,
it is expected that the saturation mechanism is insensitive to the
detailed star configuration (mass, radius, magnetic field, ...) in
which case the lower bound $\alpha_{{\rm sat}}\lesssim10^{-8}$ should
approximately hold for all sources. No saturation mechanism proposed
so far gives such a low $\alpha_{{\rm sat}}$ \cite{Lindblom:2000az,Bondarescu:2013xwa,Haskell:2013hja},
so a new mechanism would be required to make the data compatible,
via the \emph{saturated r-mode} scenario, with the interior of the
star being hadronic matter. Modifying our assumptions about the saturation
and cooling mechanisms does not qualitatively change this conclusion.
Presently proposed saturation mechanisms allow $\alpha_{{\rm sat}}$
to depend on $T$ and $\Omega$ to negative powers \cite{Alford:2012yn},
which makes the curves steeper but the intersection with the boundary
of the instability region is invariant \cite{Alford:2012yn}, so the
constraints on $\alpha_{{\rm sat}}$ are only slightly weakened. Direct
Urca cooling \cite{Yakovlev:2004iq} gives a slightly more restrictive
limit. The crust model, e.g.~with accreted envelope \cite{Potekhin:1997},
has a minor impact on the results.

We now turn to the timing data, a much larger data set of $f$ and
$\dot{f}$ for millisecond (ms) radio pulsars whose temperatures are
generally unknown. The \emph{r-mode} spindown rate is $\dot{\Omega}_{{\rm R}}\!=\!-(3\hat{G}/I)\alpha^{2}\Omega^{7}$
\cite{Owen:1998xg,Alford:2012yn}, where $I\!=\!\tilde{I}MR^{2}$
is the moment of inertia of the star. Along the thermal steady state
eq.~(\ref{eq:steady-state}) this yields the effective spindown equation
\cite{Alford:2012yn} in terms of the effective braking index $n_{{\rm RM}}\!=\!\left(\left(7\!+\!2\gamma\right)\theta\!+\!2\beta\right)/\left(\theta\!-\!2\beta\right)\leq7$.
Inverting it we find the evolution path in a $\dot{\Omega}_{{\rm R}}$-$\Omega$-plot

\begin{equation}
\Omega_{{\rm SD}}\!\left(\dot{\Omega}_{{\rm R}};\hat{\alpha}_{{\rm sat}}\right)=\left(\frac{I\hat{L}^{2\beta/\left(\theta-2\beta\right)}|\dot{\Omega}_{{\rm R}}|}{3\hat{G}^{\theta/\left(\theta-2\beta\right)}\hat{\alpha}_{{\rm sat}}^{2\theta/\left(\theta-2\beta\right)}}\right)^{1/n_{{\rm RM}}}\,.\label{eq:spindown}
\end{equation}
This equation is valid even if other spindown mechanisms---like magnetic
braking---are present, since for them the lost rotational energy does
not heat the star. By analyzing where the evolution leaves the static
instability region we obtain novel \emph{dynamic} instability boundaries
in $\dot{f}$-$f$-space. The result is in the case $\Delta\!=\!\Theta\!=\!0$

\begin{equation}
\Omega_{{\rm IB}}\!\left(\dot{\Omega}_{{\rm R}}\right)=\left(\hat{D}^{\theta}I^{\delta}|\dot{\Omega}_{{\rm R}}|^{\delta}/\left(3^{\delta}\hat{G}^{\theta}\hat{L}^{\delta}\right)\right)^{1/\left(\left(8-\psi\right)\theta-\delta\right)}\,.\label{eq:timing-instability}
\end{equation}
Note that this expression depends on the cooling behavior, but is,
like eq.~(\ref{eq:temperature-instability}), completely independent
of the saturation mechanism and amplitude. These analytic expressions
exhibit the complete dependence on the underlying physics and allow
us to make quantitative predictions with control over the uncertainties.

The timing data are plotted in the right panel of fig.~\ref{fig:exclusion-regions}.
The horizontal axis shows the amount of spindown {\em due to r-modes},
so {\em every} point is an upper limit, since other mechanisms
might contribute to the spindown. The dynamic instability boundaries
eq.~(\ref{eq:timing-instability}) are also plotted for the previously
considered star compositions %
\footnote{The envelope model \cite{Gudmundsson:1983ApJ} is used both for neutron
stars and for strange quark stars (which are assumed to have a hadronic
crust suspended by electrostatic forces \cite{Alcock:1986hz}).%
}. We see that the ungapped quark matter model of ref.~\cite{Schwenzer:2012ga}
(solid (blue) line) is compatible with the timing data, since r-modes\emph{
cannot} be the dominant spindown mechanism. As in the $T$-$f$ plot,
this is thanks to a large stability window. However, the radio data
cannot be explained by the {\em no r-mode} scenario, because once
the accretion stops theses sources quickly cool until they reach the
boundary of the instability region. If the saturated thermal steady
state curve eq.~(\ref{eq:steady-state}) is to the right of the boundary
eq.~(\ref{eq:temperature-instability}), the stars will periodically
be heated out of the instability region and cool back in again \cite{Andersson:2001ev}.
In this \emph{boundary-straddling} scenario the r-mode amplitude $\alpha_{{\rm str}}\!<\!\alpha_{{\rm sat}}$
is eventually dynamically set by thermal balance at the boundary \cite{Reisenegger:2003cq}
without the need for a non-linear saturation mechanism. We therefore
expect that non-accreting sources such as ms-pulsars will cluster
along the low-temperature boundary of the stability window in the
$T$-$f$ plane. Since there are generally no temperature estimates
for such sources we cannot yet test this prediction. However, the
timing data impose a constraint: these sources should be found on
or to the right (higher $\dot{f}$) of the stability boundary in the
$\dot{f}$-$f$ plane because other spindown mechanisms may also be
operating. This is exactly what the data (fig.~\ref{fig:exclusion-regions})
show. The interacting quark matter model is therefore compatible with
the timing data.

Concerning the hadronic matter model (long-dashed line) we first note
that the clustering of young radio sources \cite{Manchester:2004bp}
just below the hadronic matter instability boundary at $|\dot{f}|\sim10^{-10}\,{\rm s}^{-2}$
means that the hadronic model can---if $\alpha_{{\rm sat}}$ in young
pulsars is sufficiently large---explain why they do not spin faster
\cite{Alford:2012yn}. However, the large column of data points around
$|\dot{f}|\sim10^{-15}\,{\rm s}^{-2}$, consisting of old ms-pulsars,
is a problem for the minimal hadronic matter model, since these upper
limits are within its dynamic instability region. In principle this
might be explained in a saturated r-mode scenario, but as we see from
the thermal steady state curves in the right panel of fig.~\ref{fig:steady-state},
this scenario requires $\alpha_{{\rm sat}}\lesssim10^{-7}$, a similar
value to that obtained from the $T$-$f$ data, which is not provided
by current mechanisms \cite{Lindblom:2000az,Bondarescu:2013xwa,Haskell:2013hja}.

It might seem that a no-r-mode scenario is also possible, since each
point is just an upper limit, so in the $T$-$f$ plot these stars
could be at $T\!<\!10^{6}$\,K, outside the hadronic matter instability
region, so their $\dot{f}_{{\rm R}}$ is really zero. To see that
this is not possible, consider the evolutionary history of these sources:
they are recycled pulsars, previously spun up by accretion in a LMXB
\cite{Brown:1998ch}. During accretion they are heated to $T\sim10^{8}\,{\rm K}$
by nuclear reactions in the crust (left panel of fig.~\ref{fig:exclusion-regions}).
If the hadronic matter model were correct, these stars would then
become trapped in the instability region, since once the accretion
stops the stars would quickly cool until %According to the hadronic matter model, at this stage of their life%these stars are deep inside the r-mode instability region in the $T$-$f$%plot, and at a temperature where their cooling balances the heating%due to r-modes and accretion. Once the accretion stops, the stars%will quickly cool to a new steady-state temperature, where the cooling%only has to balance the r-mode heating. So in fig.~\ref{fig:steady-state}%they move to the left 
they reach the spindown line (determined by $\alpha_{{\rm sat}}$)
at which cooling balances r-mode heating. They then spin down along
the curve very slowly: at $|\dot{f}|\sim10^{-15}\,{\rm s}^{-2}$ the
frequency would change only by a few tens of Hz in a billion years.
The only way that the fastest spinning pulsars could escape the instability
region is if $\alpha_{{\rm sat}}<10^{-10}$, allowing them to cool
to $10^{5}$\,K in a few million years without crossing the r-mode
spindown line. So this scenario requires an even lower saturation
amplitude than the saturated r-mode scenario, much lower than is predicted
by any proposed r-mode saturation mechanism. This account is not contradicted
by the only ms-radio source J0437-4715, for which a temperature estimate
is available \cite{Durant:2011je}. Its spin frequency is low enough
that it could have cooled out of the instability region without crossing
an r-mode spindown curve for any $\alpha_{{\rm sat}}\lesssim10^{-6}$.
The figures also show the first LMXB (IGR J00291+5934), whose (temporary)
spindown during quiescence has been observed \cite{Patruno:2010qz}.

\begin{figure}
\includegraphics{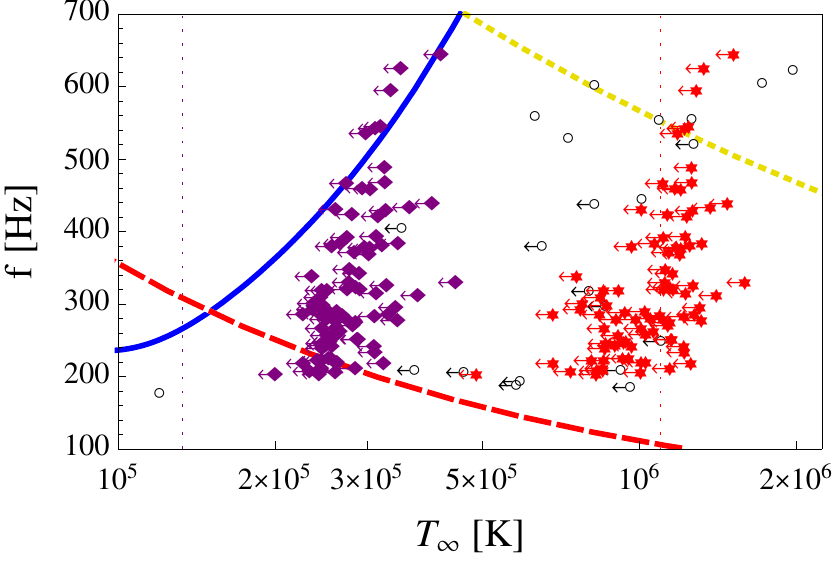}

\protect\protect\caption{\label{fig:r-mode-temperatures}The \emph{r-mode temperatures} as
observed at infinity of radio pulsars with known timing data as well
as the corresponding temperatures of LMXBs (dots) compared to r-mode
instability boundaries. R-modes temperatures are given for sources
with $f\geq200\,{\rm Hz}$ assuming neutron stars with modified Urca
(stars) and direct Urca cooling (diamonds).}
\end{figure}

Finally consider the thermal state of old radio sources. In the \emph{no
r-mode} scenario they should have cooled to very low temperatures
(potentially set by other heating sources \cite{Durant:2011je}),
whereas in the \emph{boundary straddling} scenario the temperature
would be independent of the spindown rate and determined by eq.~(\ref{eq:temperature-instability}).
Yet, if the \emph{saturated r-mode} scenario is realized then, on
each side of fig.~\ref{fig:steady-state}, ms-pulsars sit on the
spindown curve for the physical value of $\alpha_{{\rm sat}}$. If
the star is in steady state balance between r-mode heating and neutrino/photon
cooling then its temperature is a function of its spin $\Omega$ and
\emph{r-mode} spindown rate $\dot{\Omega}_{{\rm R}}$. Using eqs.~(\ref{eq:steady-state})
and (\ref{eq:spindown}) this \emph{r-mode steady-state temperature}
is 
\begin{equation}
T_{{\rm RM}}=\left(I\Omega\dot{\Omega}_{{\rm R}}/\left(3\hat{L}\right)\right)^{1/\theta}\, ,\label{eq:timing-temperature}
\end{equation}
where $I$ is the moment of inertia. The striking feature of this
simple expression is that it is \emph{independent} of the saturation
physics. This is because it is determined by rotational energy being
transformed into gravitational wave, neutrino and/or photon energy,
irrespective of the r-mode physics that accomplishes this. However,
eq.~(\ref{eq:timing-temperature}) depends on the cooling behavior
which differs for various forms of matter. The corresponding r-mode
steady-state temperatures of radio pulsars that would be observed
on earth are shown in fig.~\ref{fig:r-mode-temperatures} for the
two extreme cases of standard modified Urca cooling and fast direct
Urca cooling. If the spindown of a star is dominated by r-mode gravitational
emission then eq.~(\ref{eq:timing-temperature}) tells us its core
temperature. If only a fraction of the spindown rate is due to r-modes
then it provides only an upper bound. For the actual temperature to
be significantly below eq.~(\ref{eq:timing-temperature}) would require
that only a fraction of the observed spindown rate is due to r-modes,
which would require an even smaller value of $\alpha_{{\rm sat}}$
than the bounds obtained from fig.~\ref{fig:steady-state}, and we
do not know of any mechanism that could accomplish this. We conclude
that if radio pulsars are, as the hadronic model requires, undergoing
r-mode spindown, even tiny amplitude r-modes would have a big impact
on the thermal evolution. In this case radio pulsars should have observable
surface temperatures as measured at infinity of $\left(0.2\!-\!1\right)\!\times\!10^{6}$\,K,
which is significantly hotter than standard cooling estimates suggest
\cite{Yakovlev:2004iq}.

We conclude that the novel dynamic instability regions and the r-mode
temperature show us how timing data from the large population of radio
pulsars can constrain their interior constitution. Beyond the illustrative
examples discussed here, there are various options, like superfluid
pairing (including the important effect of mutual friction) \cite{Haskell:2012},
magnetic fields \cite{Mendell:2001nz}, hyperonic matter \cite{Reisenegger:2003cq}
or color-superconducting phases \cite{Alford:2007xm}, that might
be responsible for enhanced damping. To come to definite conclusions
will require both observational and theoretical progress. On the theoretical
side, we need saturation amplitudes, dynamic instability regions and
r-mode temperatures for all hypothesized forms of dense matter with
distinct damping and cooling properties. Observationally, it would
be particularly useful to obtain temperature measurements or bounds
for nearby ms-radio pulsars that spin with frequencies above $300\,{\rm Hz}$.
The comparison with the theoretical r-mode stability boundary could
reveal whether the \emph{saturated} or \emph{boundary-straddling}
\emph{r-mode} scenarios can be realized. If they are so cool as to
lie outside the boundary this would be inexplicable in the minimal
hadronic model. This is just one example of how the combination of
radio, x-ray and future gravitational wave data will allow us to discriminate
the \emph{no}-, \emph{saturated- or boundary-straddling r-mode} scenarios
and eventually different phases of dense matter.
\begin{acknowledgments}
We are grateful to Simin Mahmoodifar and Tod Strohmayer for helpful
discussions. This research was supported by the Offices of Nuclear
and High Energy Physics of the U.S. Department of Energy under contracts
\#DE-FG02-91ER40628, \#DE-FG02-05ER41375. 
\end{acknowledgments}
\bibliographystyle{h-physrev}
%\bibliography{cs}

\end{document}